\documentclass[showpacs,aps,prb,amsmath,amssymb]{revtex4}

\usepackage{graphicx}
\usepackage{dcolumn}
\usepackage{bm}

\begin{document}

\bibliographystyle{prsty} 

\preprint{}

\title{Extended Hubbard model with renormalized Wannier wave functions in the correlated state: beyond the parametrized models}


\author{Jan Kurzyk}
\author{W{\l}odzimierz W\'ojcik}
\affiliation{
Institute of Physics, Cracow University of Technology, Podchor\c{a}\.zych 1, 30-084 Krak\'{o}w, Poland \\
e-mail: pukurzyk@kinga.cyf-kr.edu.pl, puwojcik@kinga.cyf-kr.edu.pl
}

\author{and Jozef Spa{\l}ek}%
\affiliation{%
Marian Smoluchowski Institute of Physics, Jagiellonian University, Reymonta 4, 30 059 Krak\'{o}w, Poland \\
e-mail: ufspalek@if.uj.edu.pl \\
URL: http://th-www.if.uj.edu.pl/ztms/jspalek\_een.htm
}%
 
\begin{abstract}
The method used earlier for analysis of correlated nanoscopic systems is extended to infinite (periodic) s-band like systems described by the Hubbard model and its extensions. The optimized single-particle wave functions contained in the parameters of the Hubbard model (the hopping \textit{t} and the magnitude of the intraatomic interaction \textit{U}) are determined explicitly in the correlated state for the electronic systems of various symmetries and dimensions: Hubbard chain, square and triangular planar lattices, and the three cubic lattices (SC, BCC, FCC). In effect, the evolution of the electronic properties as a function of interatomic distance $R$ is obtained.  The model parameters in most cases do not scale linearly with the lattice spacing and hence, their solution as a function of microscopic parameters reflects only qualitatively the system evolution. Also, the atomic energy changes with $R$ and therefore should be included in the model analysis. The solutions in one dimension (\textit{D} = 1) can be analyzed both rigorously (by making use of the Lieb--Wu solution) and compared with the approximate Gutzwiller treatment. In higher dimensions (\textit{D} = 2, 3) only the latter approach is possible to implement within the scheme. The renormalized single particle wave functions are almost independent of the choice of the scheme selected to diagonalize the Hamiltonian in the Fock space in $D=1$ case. The method can be extended to other approximation schemes as stressed at the end.
\end{abstract}

\maketitle

\section{\label{sec:level_1}Introduction} 
The question of combining in an explicit manner inter-electronic correlations with single-particle (band) calculations is very important for the systems for which Coulomb interaction between electrons is comparable to or even larger than the kinetic (bare band) energy of electrons \cite{Edwards,Spalek_EurJPhys}. With respect to this, methods, starting from band calculations, such as   LDA+U \cite{Anisimov_Zaanen} or LDA+DMFT \cite{Anisimov} have been devised and they work well for quite few systems. The methods provide e.g. the photoemission spectrum  \cite{Matsura}, the overall band splitting at the Fermi surface (i.e. the stability of the Mott insulating state) optical spectrum, etc. The band structure calculations allow also for estimation of the Hubbard-interaction parameter $U$, when the Wannnier functions are determined first \cite{Schnell}. In all these methods the question of counting twice the electron-electron Coulomb extended interaction arises when the effective single-particle potential contains them.

A systematic approach base on taking into account Coulomb interactions between electrons in an exact manner first and only then determining the renormalized single-particle wave functions contained in the model parameters by a proper Euler variational procedure. In that situation we allow for an adjustment of single-particle wave functions in the correlated-electron state and only then calculate them explicitly. Such a reverse method called EDABI (\textbf{E}xact \textbf{D}iagonalization with \textbf{Ab} \textbf{I}nition \textbf{A}proach) has been devised and employed to nanoscopic chains and clusters \cite{Spalek_Podsiadly}. It provides the evolution of the correlated-system properties as a function of interatomic distance $R$, not only as a function of model parameters, which are difficult to be measure. For example \cite{Spalek_Encyclopedia}, EDABI provides new results such as e.g. a systematic evolution of the statistical distribution function as a function of increasing $R$ (from Fermi-like function to a continuous momentum distribution reflecting electron localization on parent atoms) or a magnetic Slater-like splitting of the electronic states without the appearance of long-range antiferromagnetic ordering \cite{Spalek_Podsiadly}. Needless to say that this method avoids in an explicit manner counting twice the interaction between the particles.

The purpose of this paper is to generalize and test the EDABI-type approach for extended systems of arbitrary dimension ($D=1,2,3$) described by the (parametrized) extended Hubbard model. Only in the $D=1$ case it is possible to compare an exact (Lieb--Wu, LW) solution with the approximate Gutzwiller-wave-functions (GWF), and the Gutzwiller-ansatz (GA) solutions. In higher dimensions, we calculate the single-particle properties starting from GA. What is surprising, at least for $D=1$, is the relative insensitivity of the detailed shape of the renormalized (by correlations) single-particle wave function to the method selected to diagonalize the many-particle Hamiltonian in the Fock space. In general, our method of approach completes the solution of the parametrized models in the sense that it yields the evolution of the correlated many-particle systems properties as a function of interatomic distance, as well as provides the shape of the Wannier functions in the correlated state.

The structure of the paper is as follows. In Section II and III we overview briefly our method, whereas in Section IV the extended Hubbard-chain properties are analysed in detail. In Sections V and VI selected two- and three-dimensional lattices are considered respectively. Section VII contains a brief discussion and an overview. In Appendices A and B we provide some formal details of the calculations.

\section{\label{sec:level_2} parameterized models supplemented with the single-particle basis optimization: \ \ \ \ \ \ \ \  a brief summary of EDABI}

\begin{figure*}
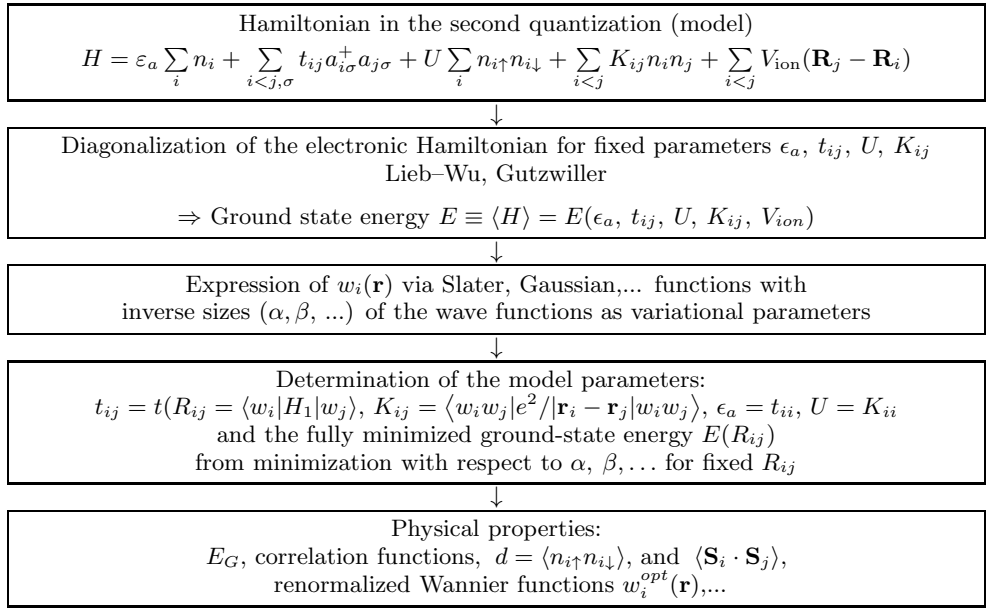

\framebox[130mm]
{
\begin{minipage}{0.8\linewidth}
\centerline{Hamiltonian in the second quantization (model)}
\vspace*{1mm}
\centerline{
$ H = \varepsilon _a \sum\limits_i {n_i }  + \sum\limits_{i<j, \sigma } t_{ij} a_{i\sigma }^ +  a_{j\sigma } + U\sum\limits_i {n_{i \uparrow } n_{i \downarrow } } 
  + \sum\limits_{i < j} {K_{ij} n_i n_j }  + \sum\limits_{i < j} {V_{{\rm{ion}}} ({\bf{R}}_j  - {\bf{R}}_i )}$
}
\end{minipage}
}

{$\downarrow$}

\framebox[130mm]
{
\begin{minipage}{0.8\linewidth}
\centerline{Diagonalization of the electronic Hamiltonian for fixed parameters $\epsilon _a,\:t_{ij},\:U,\:K_{ij}$}
\centerline{Lieb--Wu, Gutzwiller}
\vspace*{2mm}
\centerline{$\Rightarrow$ Ground state energy $E\equiv\left\langle H \right\rangle =E(\epsilon_a,\:t_{ij},\:U,\:K_{ij},\:V_{ion})$}
\end{minipage}
}

{$\downarrow$}

\framebox[130mm]
{
\begin{minipage}{0.8\linewidth}
\centerline{Expression of $w_i(\textbf{r})$ via Slater, Gaussian,... functions with} 
\centerline{inverse sizes ($\alpha, \beta,$ ...) of the wave functions as variational parameters}
\end{minipage}
}

{$\downarrow$}

\framebox[130mm]
{
\begin{minipage}{0.8\linewidth}
\centerline{Determination of the model parameters: } 
\centerline{$t_{ij}=t(R_{ij}=\left\langle w_i|H_1|w_j\right\rangle$, $K_{ij}=\left\langle w_iw_j|e^2/|\textbf{r}_i-\textbf{r}_j|w_iw_j\right\rangle$, $\epsilon_a=t_{ii}$, $U=K_{ii}$}
\centerline{and the fully minimized ground-state energy $E(R_{ij})$}
\centerline{from minimization with respect to $\alpha,\:\beta, \ldots$ for fixed $R_{ij}$}
\end{minipage}
}

{$\downarrow$}

\framebox[130mm]
{
\begin{minipage}{0.8\linewidth}
\centerline{Physical properties:}
\centerline{ $E_G,$ correlation functions, $\:d=\left\langle n_{i\uparrow}n_{i\downarrow}\right\rangle$, and $\:\left\langle \textbf{S}_i\cdot\textbf{S}_j\right\rangle$, }
\centerline{ renormalized Wannier functions $w_i ^{opt}(\textbf{r})$,...}
\end{minipage}
}
\caption{\label{fig:flowchart} Flowchart of the approach combining exact diagonalization in the \textit{Fock space} followed by the single-particle wave-function optimization in the \textit{Hilbert space} for a correlated state. For details see main text.}
\end{figure*}

Before placing our work in the literature of the subject, we first characterize the essence of our approach \cite{Spalek_Podsiadly} from the formal side. The method of the optimized single-particle wave functions incorporates the first and the second quantization schemes. Namely, if $\Psi(\textbf{r}_1,...,\textbf{r}_N)$ describes the \textit{N}-particle wave function in the Schr\"odinger representation (i.e. in \textit{Hilbert space}), then in the second-quantization representation (i.e. in the \textit{Fock space}) this state can be represented by \cite{Schweber}
\begin{equation}
\left| {\left. \Phi  \right\rangle } \right. = \frac{1}{{\sqrt {N!}} }\int {d^3 {\bf{r}}_1...d^3 {\bf{r}}_N  \Psi ({\bf{r}}_1 ,...,{\bf{r}}_N )\hat \Psi ^ \dagger  ({\bf{r}}_1 )...\hat \Psi ^ \dagger  ({\bf{r}}_N )}  \left| {\left. 0 \right\rangle }, \right.\label{eq:1}
\end{equation}
where, $\hat \Psi ^ \dagger(\bf{r}_i) \equiv \left(\hat \Psi _{\uparrow} ^ \dagger (\bf{r}_i) \hat \Psi _{\downarrow} ^ \dagger (\bf{r}_i) \right)$, denotes the field operators representing the particle creation at point $\bf{r}_i$ with spin $\sigma=\uparrow,\downarrow\equiv\pm 1$. Utilizing the anticommutation relations for the field operators one can easily find the inverse representation which has the form

\begin{equation}
\Psi ({\bf{r}}_1 ,...,{\bf{r}}_N ) = \frac{1}{{\sqrt {N!} }}\left\langle 0 \right|\hat \Psi ({\bf{r}}_1 )...\hat \Psi (\textbf{r}_N )\left| \Phi  \right\rangle\label{eq:2}.
\end{equation}

So, the two schemes are equivalent if only the anticommutation relations of the field operators are defined. Here, the field operator is defined in the single-particle basis of Wannier functions ${w_i(\textbf{r})}$ as follows

\begin{equation}
\hat \Psi({\bf{r}}) = \sum\limits_{i=1, \sigma=\pm1}^{\infty} {w_i } \chi_\sigma ({\bf{r}})a_{i\sigma }^{}\equiv \sum\limits_{i=1}^{\infty} {w_i }({\bf{r}})\left( \begin{array}{c}a_{i\uparrow}\\a_{i\downarrow} \end{array} \right) ,
\label{eq:3}
\end{equation}
where the summation over \textit{i} runs over a complete basis of single-particle states. The basis ${\{w_i(\bf{r})}\}$ can be arbitrary, provided it is complete in the quantum-mechanical sense. In most situations we introduce \textit{a model}, i.e. select the subset of a complete basis $\{i\}$. For example we replace this complete basis by a \textit{finite} subset ${\{w_i(\bf{r})}\}, i=1,...,M$ of the functions, connected directly to the problem at hand. Hence, in the single narrow-band situation we have that \cite{Spalek_Podsiadly}

\begin{equation}
\hat \Psi ({\bf{r}}) \approx \sum\limits_{i=1}^{M} {w_i } ({\bf{r}})a_i^{},
\label{eq:4}
\end{equation}
where in what follows we set $M=N$ and $\{w_i(\bf{r})\}$ is the single-particle-Wannier-function basis (to be defined later). In effect, the approximate states in the Fock space are defined through

\begin{equation}
\left| \Phi  \right\rangle  \approx \frac{1}{\sqrt {N!}}\sum\limits_{j_1 ...j_N  = 1}^M {C_{j_1 ...j_N } } a_{j_1 }^ +  ...a_{j_N }^ +  \left| 0 \right\rangle ,
\label{eq:5}
\end{equation}
and the $N$-particle wave function in this approximate basis has now the form

\begin{equation}
\Psi ({\bf{r}}_1 ,...,{\bf{r}}_N ) = \frac{1}{{N!}}\sum\limits_{i_1...i_N }^M {\sum\limits_{j_1...j_N }^M {\left\langle 0 \right|a_{i_N } ...a_{i_1 } a_{j_1 }^ +  ...a_{j_N }^ +  \left| 0 \right\rangle } }
\times C_{j_1 ...j_N } w_{i_1 } ({\bf{r}}_1 )...w_{i_N } ({\bf{r}}_N ).
\label{eq:6}
\end{equation}

The coefficients $C_{j_1 ...j_N }$ can be calculated by either direct Hamiltonian diagonalization or by Lanczos method for finite systems \cite{RycerzPhD}, whereas the normalized (optimized) wave functions ${\{w_i(\bf{r})\}}$ are determined by the procedure described below. Note that the second-quantization formalism separates the many-particle function aspect of the problem, which is contained in the coefficients $C_{j_1...j_N}$, from the wave-mechanics aspect of determining the basis $\{w_i(\bf{r})\}$. In fact, the two are intertwinned. Namely, we perform the diagonalization of the second-quantized Hamiltonian for selected (and fixed) single-particle basis first and only then optimize the basis $\{w_i(\bf{r})\}$ with the help of a variational approach.

We now describe how to combine the second and the first quantization schemes on the example of extended Hubbard model. The general Hamiltonian for interacting particles is

\begin{equation}
H = \sum\limits_{ij\sigma } {t_{ij} {a_{i\sigma }^ +  a_{j\sigma } } + \frac{1}{2}\sum\limits_{ijkl\sigma \sigma '} {V_{ijkl} } {a_{i\sigma }^ +  a_{j\sigma '}^ +  a_{l\sigma '} a_{k\sigma } } } ,
\label{eq:7}
\end{equation}
where

\begin{equation}
t_{ij}  \equiv \left\langle {w_i } \right|H_1 \left| {w_j } \right\rangle \equiv \int {d^3 {\bf{r}}} w_i^* ({\bf{r}})H_1 w_j ({\bf{r}}),
\label{eq:8}
\end{equation}
is the hopping integral with

\begin{equation}
H_1({\bf{r}})  =  - \frac{{\hbar ^2 }}{{2m}}\nabla ^2  - \sum\limits_j {\frac{{e^2 }}{{|{\bf{r}} - {\bf{R}}_j |}}}  \mathop \equiv\limits^{a.u.}-\frac{1}{2}\nabla^2-\sum\limits_j{\frac{{2}}{{|{\bf{r}} - {\bf{R}}_j |}}}
\label{eq:9} 
\end{equation}
being the Hamiltonian of single \textit{bare} electron (a.u.=atomic units), and is the Hamiltonian for single electron in the system under consideration, and

\begin{equation}
V_{ijkl}  \equiv \left\langle {w_i w_j } \right|V\left| {w_k w_l } \right\rangle  =
\int {d^3 {\bf{r}}_1 d^3 {\bf{r}}_2 w_i^* ({\bf{r}}_1 )w_j^* ({\bf{r}}_2 )} V({\bf{r}}_1  - {\bf{r}}_2 )w_k ({\bf{r}}_1 )w_l ({\bf{r}}_2 ) 
\label{eq:10}
\end{equation}
is the amplitude of classical Coulomb interaction rewritten in the first-quantization language. Note that here $i$ labels complete set of quantum numbers except spin. The ground state energy is then determined by the expression

\begin{equation}
E_G  \equiv \left\langle H \right\rangle  = \sum\limits_{ij\sigma } {t_{ij} \left\langle {a_{i\sigma }^\dagger  a_{j\sigma } } \right\rangle } + \frac{1}{2}\sum\limits_{ijkl\sigma \sigma '} {V_{ijkl}  \left\langle {a_{i\sigma }^ +  a_{j\sigma '}^ +  a_{l\sigma '} a_{k\sigma } } \right\rangle },
\label{eq:11}
\end{equation}
where the ground-state averages $\left\langle ... \right\rangle $ are determined for the many particle ground state $\left| \Phi  \right\rangle \equiv \left| \Phi_G \right\rangle$ for fixed $\{w_i(\bf{r})\}$. This means, that the trial basis $\{w_i(\bf{r})\}$ $i=1,...,M$ entering $t_{ij}$ and $V_{ijkl}$ must be optimized, i.e. \textit{the ground state energy must be a minimum within the class of trial (incomplete) basis of wave-functions}. Obviously, such an optimization would not be necessary if the basis were complete. In the present approach this means that we have to construct the functional variational scheme to determine the optimized trial basis $\{w_i(\bf{r})\}$. This formal procedure provides us with the evolution of the system properties as a function of the lattice parameter \textit{R}. The method is summarized schematically in Fig.~\ref{fig:flowchart}. 

\section{\label{sec:level_3}The model}
\subsection{\label{sec:level_3.1}Extended Hubbard model}

To describe a single-band model of interacting fermions we start with the extended Hubbard Hamiltonian

\begin{equation}
 H = \varepsilon _a \sum\limits_i {n_i }  + \sum\limits_{i<j, \sigma } t_{ij} a_{i\sigma }^ +  a_{j\sigma } + U\sum\limits_i {n_{i \uparrow } n_{i \downarrow } } 
  + \sum\limits_{i < j} {K_{ij} n_i n_j }  + \sum\limits_{i < j} {V_{{\rm{ion}}} ({\bf{R}}_j  - {\bf{R}}_i )}, \label{eq:12}
\end{equation}
where now $i$ labels a Wannier orbital centered at $i\equiv\textbf{R}_i$, $\varepsilon _a \equiv t_{ii}$, is the atomic energy per site, $t_{ij}$ - the hopping integral between the sites $i$ and $j\neq i$, $U$ is the magnitude of intrasite Coulomb interaction, whereas $K_{ij}$ is the corresponding quantity for electrons located on sites $i$ and $j$ with $j\neq i$. Finally,

\begin{equation}
V_{{\rm{ion-ion}}} \mathop  = \limits^{a.u.} \frac{2}{{|{\bf{R}}_i  - {\bf{R}}_j |}} = \frac{2}{{R_{ij} }}
\label{eq:13}
\end{equation}
is the classical Coulomb interactions between the cations located at the sites $i$ and $j$ in atomic units (a.u.). One should note that above $i$ and $j$ mean the ionic positions, so the Hubbard model already does not base on a \textit{complete set} of the single-particle wave-functions $\{w_i(\bf{r})\}$, i.e. neglects $p,d,f,$ etc. Wannier states.

Even though we study here the Hubbard model, we have to include the intersite Coulomb interaction if the atomic limit is to be recovered properly in the limit of interatomic distance $R_{ij}\rightarrow\infty$. In order to achieve that, we first represent the intersite term in the form ~\cite{RycerzPhD}

\begin{widetext}
\begin{eqnarray}
\sum\limits_{i < j} {K_{ij} n_i n_j } = \sum\limits_{i < j} {K_{ij} } (n_i  - 1)(n_j  - 1)
- \sum\limits_{i<j} {K_{ij}  + 2N_e \frac{1}{N}\sum\limits_{i < j} {K_{ij} } }
\nonumber\\ 
= \sum\limits_{i < j} {K_{ij} } \delta n_i \delta n_j  + N_e \frac{1}{N}\sum\limits_{i < j} {K_{ij} }  + (N_e  - N)\frac{1}{N}\sum\limits_{i < j} {K_{ij} }\label{eq:14} ,
\end{eqnarray}
\end{widetext}
where $N_e\equiv \sum \limits_{i} {n_i}$ is the total number of electrons, $N$ is the number of lattice sites, and $\delta n _i\equiv n _i-1$. In the Mott insulating state the number of electrons $N_e=N$ i.e. in the half-filled case, $\langle \delta n_i \rangle =0$. This condition defines the Mott-Hubbard state. In this state, we have that

\begin{equation}
\sum\limits_{i < j} {K_{ij} n_i n_j  = \sum\limits_{i < j} {K_{ij} } }. 
\label{eq:15}
\end{equation}
In effect, Hamiltonian ~(\ref{eq:12}) in that limit assumes the form
\begin{equation}
H = \varepsilon _a^{{\rm{eff}}} \sum\limits_i {n_i }  + \sum\limits_{i<j, \sigma } {t_{ij} a_{i\sigma }^ +  a_{j\sigma }}  + U\sum\limits_i {n_{i \uparrow } n_{i \downarrow } },
\label{eq:16}
\end{equation}
where
\begin{equation}
\varepsilon _a^{e{\rm{ff}}}  \equiv \varepsilon _a  + \frac{1}{N}\sum\limits_{i < j} {\left( {K_{ij}  + \frac{2}{{R_{ij} }}} \right)},
\label{eq:17}
\end{equation}
is the effective atomic energy which reduces to the true atomic energy in the limit of large interatomic separation. The inclusion of this energy part assure the reduction of $\varepsilon^{eff}_{a}$ to that of the isolated atoms for $R_{ij}\rightarrow\infty$. In that situation, $\varepsilon^{eff}_{a}$ is not a constant quantity, as would be the case for a parameterized model. The ground state energy is

\begin{equation}
\frac{{E_G }}{N} = \varepsilon _a^{{\rm{eff}}} + \frac{1}{N}\left( {\sum\limits_{i<j, \sigma } {t_{ij}  \langle a_{i\sigma }^ +  a_{j\sigma }  \rangle }  + U\sum\limits_{} { \langle n_{i \uparrow } n_{i \downarrow }  \rangle } } \right).
 \label{eq:18}
\end{equation}

One sees that in this expression we have neglected the contribution of charge fluctuations $\sim\sum K_{ij}\left\langle \delta n_i \delta n_j\right\rangle$. Obviously, it vanishes in the Hartree-Fock approxiamtion. It constitutes the only approximation, under which the Lieb--Wu solution can be applied to the extended Hubbard chain as well.

\subsection{\label{sec:level_3.2}Constructions of the single-particle basis and the construction of the attractive periodic potential}

We first construct the single-particle wave-function basis $\{w_i(\textbf{r})\}$ entering the expressions for $t$, $U$, and $K_{ij}$ in the tight binding approximation i.e. assume that
\begin{equation}
w_i ({\bf{r}}{\rm{)}} \equiv \beta \Psi _i ({\bf{r}}{\rm{)}} - \gamma \sum\limits_{j = 1}^z {\Psi _j } ({\bf{r}}), \label{eq:20}
\end{equation}
where $z$ is the number of nearest neighbors, $\beta$ and $\gamma$ are the mixing coefficients, and

\begin{equation}
\Psi _i ({\bf{r}}) \equiv \sqrt {\frac{{\alpha ^3 }}{\pi }} e^{ - \alpha |{\bf{r}} - {\bf{R}}_i |}
\label{eq:21}
\end{equation}
is the $1s$ atomic wave function centered on the site $i$. In concrete calculations, they are represented by the Gaussians, as defined below. The parameters $\beta$ and $\gamma$ are selected to fulfill the atomic--limit properties

\begin{subequations}
\label{eq:22}
\begin{equation}
\mathop {\lim }\limits_{R \to \infty } \alpha  = \alpha _0  = 1/a_0 ,
\end{equation}
where $a_0$ is the atomic Bohr radius and
\begin{equation}
\mathop {\lim }\limits_{R \to \infty } \beta  = 1,{\rm{     }}\mathop {\lim }\limits_{R \to \infty } \gamma  = 0.
\end{equation}
\end{subequations}
As the functions $\{w_i(\textbf{r})\}$ are orthogonalized atomic orbitals, we have that
\begin{equation}
 < w_i |w_i  >  = 1,\ \ \ \ < w_i |w_j  >  = 0.
\label{eq:23}
\end{equation}
This conditions lead to the explicit expressions of the form
\begin{equation}
\beta  = \frac{{A + \sqrt {A^2  - BS_1 } }}{{[2A^2  - BS_1  - zAS_1^2  + 2(A - zS_1^2 )\sqrt {A^2  - BS_1 } ]^{1/2} }},
\label{eq:24}
\end{equation}
and
\begin{equation}
\gamma  = \frac{{S_1 }}{{[2A^2  - BS_1  - zAS_1^2  + 2(A - zS_1^2 )\sqrt {A^2  - BS_1 } ]^{1/2} }},
\label{eq:25}
\end{equation}
with
\begin{equation}
A \equiv \sum\limits_{j_1 (i) = 1}^z { \langle \Psi _j |\Psi _{j_1 (i)}  \rangle }  = \sum\limits_{j_1 (j) = 1}^z { \langle \Psi _i |\Psi _{j_1 (j)}  \rangle },
\label{eq:26}
\end{equation}
and
\begin{equation}
B \equiv \sum\limits_{j_1 (i),j_2 (j) = 1}^z { \langle \Psi _{j_1 (i)} |\Psi _{j_2 (j)}  \rangle },
\label{eq:27}
\end{equation}
whereas
\begin{equation}
S_1  \equiv  \langle \Psi _{{\bf{R}}_i } |\Psi _{{\bf{R}}_{i + 1} }  \rangle
\label{eq:28}
\end{equation}
is the overlap integral. In general,
\begin{equation}
S_n  \equiv  \langle \Psi _{{\bf{R}}_i } |\Psi _{{\bf{R}}_{i+n} }  \rangle
\label{eq:29}
\end{equation}
is the overlap that with the $n$-th coordination sphere. The symbol $j(i)$ in the above equations labels $j$ neighbors of the site $i$. The explicit form of the coefficients $A$ and $B$ for the structures studied is provided in Appendix A.

In the BCC structure $S_1$ represents overlap between the neighbors at distance $(\sqrt{3}/2)R$, whereas for FCC it is $(\sqrt{2}/2)R$, where $R$ is the lattice parameters. We use these expressions to determine the self-adjusted wave functions $\{w_i(\textbf{r})\}$.

\begin{figure*}
\centerline{
\includegraphics[width=14cm]{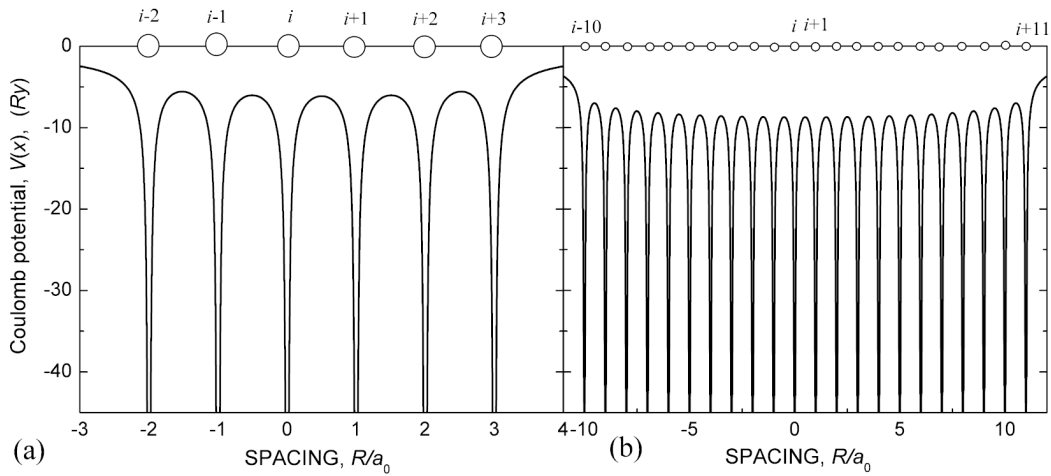}}
\caption{\label{fig:wells} Effective "periodic" attractive Coulomb potential seen by electron localized on site $i$ of a linear chain of $(+e)$ kations and composed of: (a) 6 wells $(j \in   S_2(i)\cup S_2(i+1))$ and (b) 22 wells $(j \in S_{10}(i)\cup S_{10}(i+1))$. For details see main text.}
\end{figure*}

To proceed with the numerical calculations a proper choice of the single-particle attractive potential in Eq.~(\ref{eq:9}) has to be made which expressed the periodic kation potential energy. For that purpose, we have selected either 20 or 6 potential wells surrounding given central ion, as shown in Fig.~\ref{fig:wells}.

For such defined single-particle basis, the parameters $\varepsilon _a$ and $t$ acquire the form for a linear chain 
\begin{subequations}
\begin{eqnarray}
\varepsilon _a = \beta ^2 T_0  - 4\beta \gamma T_1  + 2\gamma ^2 (T_0  + T_2 ),\ \ \ \ \ \ \ \ \ \ \ \ \ \ \ \ \ \\
-t \equiv t_{i,i + 1} = \beta ^2 T_1  - 2\beta \gamma (T_0  + T_2 ) + \gamma ^2 (3T_1  + T_3 ),\ \ \  \ \ \ \label{eq:30}
\end{eqnarray}
\end{subequations}
where
\begin{equation}
T_k  \equiv  \langle \Psi _i |H_1 |\Psi _j  \rangle  \ \ \ \ \text{for} \ \ \ j \in S_k (i),
\label{eq:31}
\end{equation}
are the hopping integrals in the starting (Slater or Gaussian) basis. The corresponding quantities for the remaining lattices are provided in Appendix B.

\begin{figure}
\centerline{
\includegraphics[width=10.5 cm]{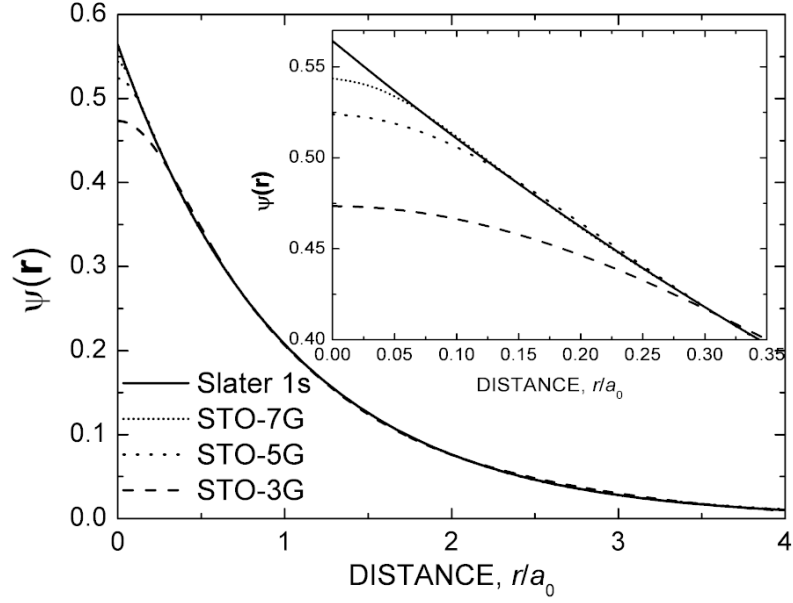}}
\caption{\label{fig:STO-nG}Comparison of $1s$ Slater wave function (solid line) with its STO-3G, STO-5G and STO-7G representations. Inset: the difference in values near $|\textbf{r}|=0$.}
\end{figure}

The corresponding expressions for $U$ and $K_{ij}$ are more involved and will not be reproduced here. As the interaction parameters involve 6-fold integrals of the product of four wave functions, we have decided to use the Gaussian representation of $\Psi _i(\textbf{r})$ of adjustable size rather than the Slater orbitals (\ref{eq:20}) \cite{gauss1,gauss2,gauss3}. Namely, we shall use STO-nG basis. The coefficients of Gaussians are determined from their best adjustment to the $1s$ Slater wave function. In Fig.~\ref{fig:STO-nG} we exhibit the hydrogen $1s$ state representation in representations STO-3G, STO-5G and STO-7G, for which the ground state energy is respectively equal to $-0.99169\:Ry$, $-0.99912\:Ry$ and $-0.99987\:Ry$. Important is that the tail part playing the most important role in the narrow-band limit is reproduced quite well in the all cases. Also, one has to remember that the interaction parameters have a relatively large interatomic part. 

\section{\label{sec:level_4}Lieb--Wu solution combined with Ab Initio adjustment of wannier functions}
\subsection{\label{sec:level_4.1}Ground state properties}

The Hubbard-chain case $(D=1)$ with the exact Lieb--Wu (LW) diagonalization provides the canonical example of  the application of EDABI method to the infinite systems. Here we treat also the LW solution as the only rigorous test of our approach.

We start with the Lieb--Wu expressions including additionally $\varepsilon _a^{eff}$, as we will study the system evolution as a function of interatomic distance (not only as a function of model parameters, as they are varying with $R$). Explicitly, the ground-state energy expression is then \cite{LiebWu}
\begin{equation}
\frac{{E_G }}{N} = \varepsilon _a^{eff}  - 4t\int\limits_0^\infty  {\frac{{J_0 (\omega )J_1 (\omega )}}{{\omega \left( {1 + e^{\omega U/(4t)} } \right)}}} d\omega , 
\label{eq:32}
\end{equation}
where $J_n(x)$ is the $n$-th order Bessel function. For the sake of comparison, the corresponding expression for the ground-state energy in the GWF approximation takes the form
\begin{equation}
\frac{{E_G }}{N} = \varepsilon _a^{eff}  - 4t\int\limits_{ - \pi }^\pi  {dk\cos (k)n_k (g)}  + Ud(g),
\label{eq:33}
\end{equation}
where $g$ is the Gutzwiller variational parameter, $n_k$ is the momentum distribution (see \cite{Metzner_Vollhardt1, Metzner_Vollhardt2}) and $d$ expresses the probability of double occupancy. Also, the corresponding analytical expression in the Gutzwiller ansatz (GA) is
\begin{equation}
\frac{{E_G }}{N} = \varepsilon _a^{eff}  - \frac{{4t}}{\pi }\left( {1 - \frac{{\pi U/t}}{{32}}} \right)^2. 
\label{eq:34}
\end{equation}
These formulas represent a final step in solving model in LW, GWF and GA schemes, respectively. Here, these expressions represent a starting point for optimization of the single-particle basis $\{w_i(\textbf{r})\}$. We proceed as follows. First, we fix the interatomic distance $R$ and for that distance construct the Wannier basis using the form of the periodic potential shown in Fig.~\ref{fig:wells} and the form (\ref{eq:20}) of the tight binding wave function. Then, we utilize the expression (\ref{eq:32}) to calculate $E_G$ for given $\varepsilon _a^{eff}$, $t$, $U$, which depend on the value of $R$. To be able to calculate the parameters $U$, $K_{ij}$ (as well as $t$ and $\varepsilon _a^{eff}$ in a single approach), we utilize the Gaussian basis STO-3G and STO-7G. In effect, the exact Lieb--Wu expression (\ref{eq:32}) turns into a function, which is minimized with respect to $\alpha$. This means that we change $\alpha \rightarrow \alpha + \delta\alpha$ and calculated $E_G$ again and repeat the whole procedure until we reach the $E_G(\alpha)$ minimum for given $R$. This minimal energy represents the true physical ground-state energy and the corresponding value of $\alpha=\alpha_{min}$ as the inverse size of atomic wave-function in the correlated state.

\begin{figure}
\centerline{
\includegraphics[width=10.5cm]{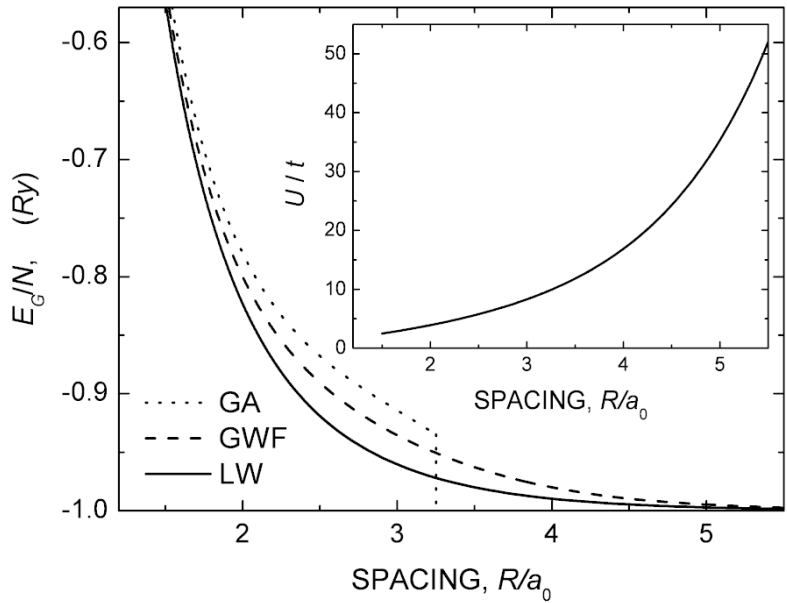}}
\caption{\label{fig:EG_1D}Ground state energy for the correlated chain vs. interatomic distance $R$ (in units of Bohr radius $a_0$) for the three methods of solving the extended Hubbard model, discussed in main text. Inset: $R$ dependence of $U/t$ ratio.}
\end{figure}

In Fig.~\ref{fig:EG_1D} we display the ground state energy as a function of interatomic distance for LW, GWF and GA methods of approach.

\begin{figure}
\centerline{
\includegraphics[width=10.5cm]{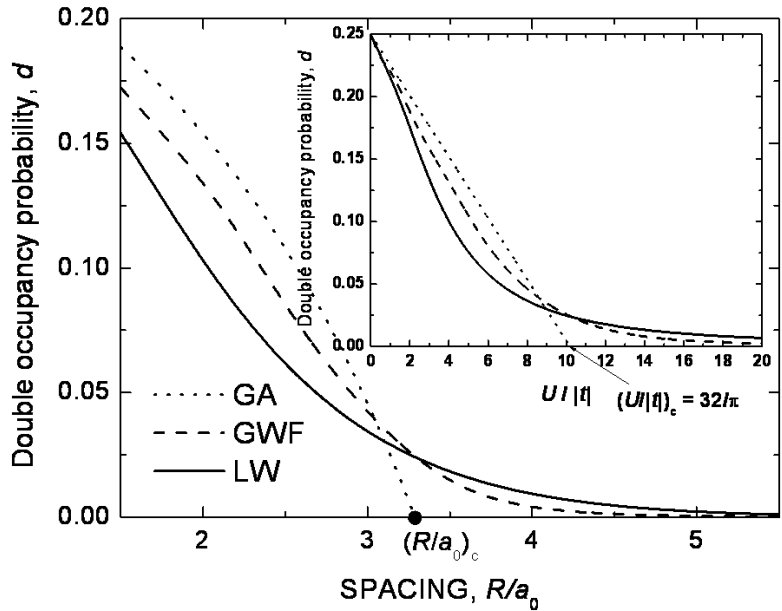}}
\caption{\label{fig:d_1D}Double occupancy probability $d= \langle n_{i\uparrow}n_{i\downarrow}\rangle$ vs. $R$  for the three solutions, using STO-7G basis. Inset: Double occupancy probability vs. $U/t$. The critical value of $(R/a_0)_c$, corresponding to the spurious Mott localization in GA, is marked on x axis. Inset: $d$ vs. $U/t$.}
\end{figure}

As one can expect, the exact (LW) solution provides the lowest energy for all values of $R$. However, the energy is higher than that of separated hydrogen atoms. So, the (extended) Hubbard chain with $1s$ states is not stable in the vacuum. Such a quantum wire can be thus created only on an appropriate substrate.
The atomic limit is practically reached for $R \agt 5.5\:a_0$, which corresponds to $U/t>50$ (see the inset in Fig.~\ref{fig:EG_1D}). For GA method the function $E_G(R)$ ends at $R\approx3.3\:a_0 \approx 1.7$ \AA \, for an optimized Wannier function with $\alpha \approx 1.021 a_0$, $\beta \approx 1.067$, and $\gamma \approx 0.157$. The important feature of the solution is that the $U/t$ dependence on $R$ is nonlinear, and thus those two representations of the results (either as a function of $U/t$ or vs. $R$) are not equivalent. It is also important to note that in the inset in Fig.~\ref{fig:EG_1D} we show $U/t$ for \textit{all} methods discussed; the curves practically coincide. Similar coincidence takes place also in the case of the parameters $\varepsilon_a^{eff}$, $t$, $U$, $K\equiv K_{i,i+1}$, and related quantities, as shown in Table~\ref{tab:tab_LW_GWF_GA}. The same conclusion holds even when STO-3G basis is utilized. One can see also that the atomic value of $U=(5/4)$ Ry is reached for $R\simeq 6\:a_0$. However, the three methods provide quite different values of double occupancy probability $d$, as displayed in Fig.~\ref{fig:d_1D}.

\begin{table}
\caption{\label{tab:tab_LW_GWF_GA}Microscopic parameters for the Hubbard chain as a function of lattice spacing for the three methods utilized. The STO-7G basis was used. For details see main text.}
\begin{tabular}{c|ccc||ccc||ccc||ccc}
\hline \hline 
  & LW & GWF & GA & LW & GWF & GA & LW & GWF & GA & LW & GWF & GA\\
\cline{2-13}  
$R/a_0$&&$\varepsilon _a^{eff}$ (Ry)& & & $-t$ (Ry)& & &$U$ (Ry)& & &$K$ (Ry)\\
\hline
1.5&0.055&0.054&0.053&-0.814&-0.813&-0.811&2.033&2.031&2.029&1.165&1.165&1.164\\
2.0&-0.568&-0.569&-0.570&-0.438&-0.437&-0.435&1.712&1.708&1.703&0.910&0.909&0.908\\
2.5&-0.804&-0.805&-0.806&-0.265&-0.262&-0.261&1.527&1.517&1.510&0.750&0.749&0.748\\
3.0&-0.906&-0.907&-0.907&-0.171&-0.169&-0.168&1.416&1.404&1.394&0.640&0.639&0.638\\
3.5&-0.954&-0.954&-0.938\footnotemark[1]&-0.114&-0.114&-0.133\footnotemark[1] &1.348&1.341&1.353\footnotemark[1]&0.557&0.557&0.588\footnotemark[1]\\
\hline
4.0&-0.977&-0.977&&-0.078&-0.077&&1.308&1.305&&0.493&0.493&\\
5.0&-0.994&-0.994&&-0.036&-0.036&&1.268&1.269&&0.399&0.399&\\
6.0&-0.999&-0.999&&-0.016&-0.016&&1.255&1.255&&0.333&0.333&\\
7.0&-1.000&-1.000&&-0.007&-0.007&&1.251&1.251&&0.286&0.286&\\
8.0&-1.000&-1.000&&-0.003&-0.003&&1.250&1.250&&0.250&0.250&\\
$\infty$&-1&-1&&0&0&&1.25&1.25&&0&0&\\
\hline \hline
$R/a_0$&& $\beta$ & & & $\gamma$ & & & $\alpha\:a_0$ & & & $S$\\
\hline
1.5&1.435&1.437&1.440&0.472&0.473&0.475&1.332&1.331&1.328&0.587&0.588&0.589\\
2.0&1.256&1.259&1.262&0.337&0.339&0.342&1.182&1.178&1.174&0.491&0.493&0.495\\
2.5&1.151&1.156&1.159&0.246&0.250&0.254&1.100&1.091&1.084&0.401&0.406&0.410\\
3.0&1.087&1.091&1.094&0.180&0.184&0.188&1.055&1.044&1.034&0.317&0.323&0.328\\
3.5&1.048&1.050&1.067\footnotemark[1]&0.131&0.133&0.157\footnotemark[1]&1.031&1.023&1.021\footnotemark[1]&0.243&0.247&0.283\footnotemark[1]\\
\hline
4.0&1.026&1.026&&0.095&0.095&&1.018&1.015&&0.181&0.183&\\
5.0&1.007&1.007&&0.048&0.048&&1.006&1.006&&0.095&0.095&\\
6.0&1.002&1.002&&0.023&0.023&&1.002&1.002&&0.046&0.046&\\
7.0&1.000&1.000&&0.010&0.010&&1.000&1.001&&0.021&0.021&\\
8.0&1.000&1.000&&0.004&0.004&&1.000&1.000&&0.009&0.009&\\
$\infty$&1&1&&0&0&&1&1&&0&0&\\
\hline \hline
\end{tabular}
\footnotetext[1]{The value for $(R/a_0)_c \approx 3.288$ is the threshold for the spurious metal-insulator transition within GA method.}
\end{table}

\subsection{\label{sec:level_4.2}Discussion of the results for the Hubbard chain}

\begin{figure}
\centerline{
\includegraphics[width=10.5cm]{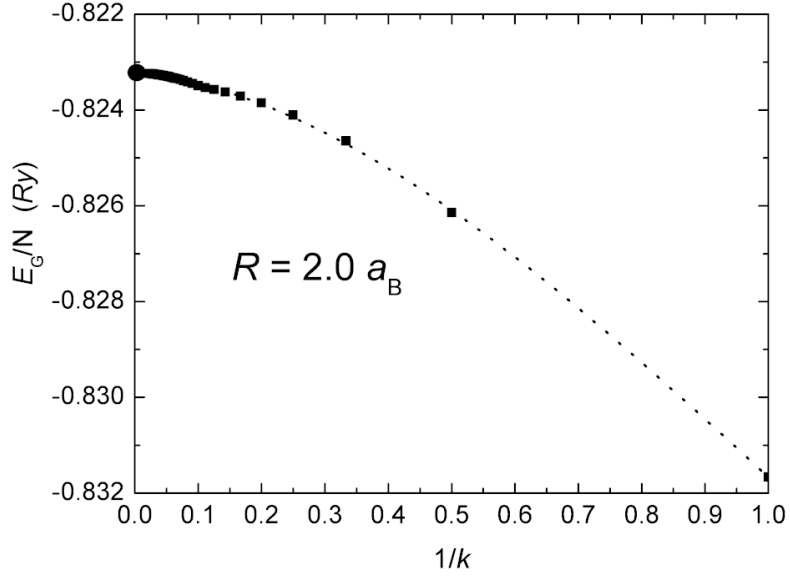}}
\caption{\label{fig:EG_vs_k} Ground state energy of the infinite Hubbard chain for the lattice parameter $R=2a_0$ ($a_0$ - Bohr radius) obtained starting from the Lieb--Wu solution, plotted as a function of inverse $k$, where $k$ denotes the number of coordination spheres taken to represent the potential $V(\textbf{r})$. The limit $1/k \rightarrow 0$ represents the exact value for this lattice in the tight-binding approximation with 1s orbitals.}
\end{figure}

\begin{figure}
\centerline{
\includegraphics[width=10.5cm]{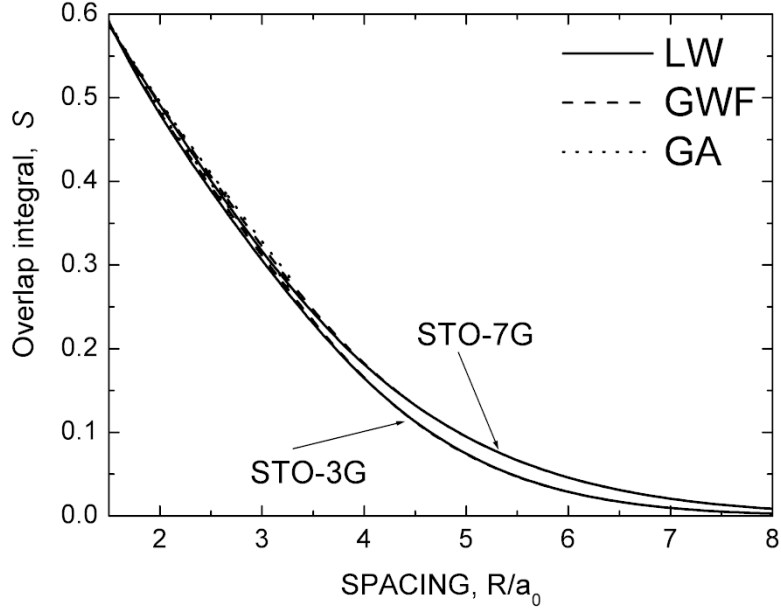}}
\caption{\label{fig:S_vs_R_CH} Overlap integral $S_1$ vs. interatomic distance, calculated for optimized the Wannier functions within the three schemes (LW, GWF, GA) and for the two selections of bases respectively.}
\end{figure}

\begin{figure}
\centerline{
\includegraphics[width=10.5cm]{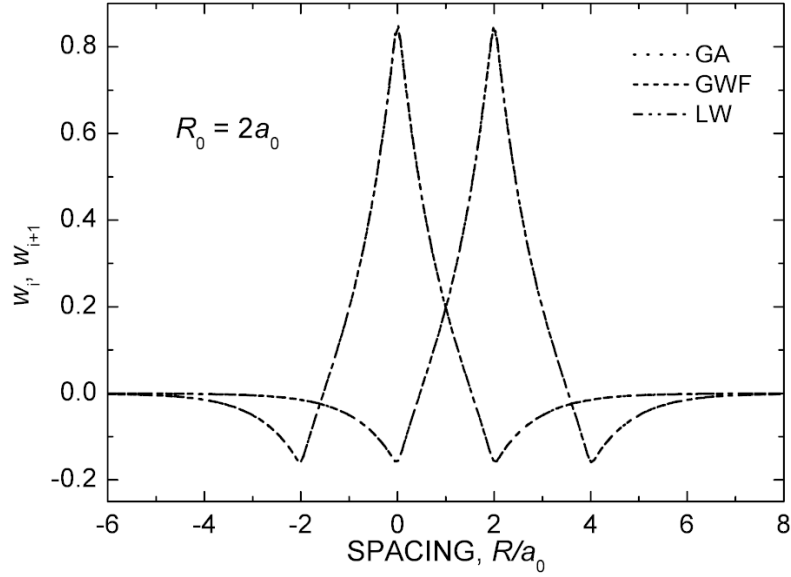}}
\caption{\label{fig:wanniers}The example of the two nearest neighboring optimized Wannier functions for the Hubbard chain in the tight-binding approximation and using the STO-7G starting Gaussian basis, within the three schemes (LW, GWF, GA). The parameters of the Wannier functions are as follows: $\alpha=1.194$, $\beta=1.256$, $\gamma=0.341$ (LW); $\alpha=1.199, \beta=1.252$, $\gamma=0.338$ (GWF); and $\alpha=1.204$, $\beta=1.248$, $\gamma-0.335$ (GA).}
\end{figure}

We first discuss the question to what extent the assumptions influence the numerical results. The first of them is the choice of the number of potential wells (cf. Fig.~\ref{fig:wells}). In Fig.~\ref{fig:EG_vs_k} the ground state energy for LW solution using STO-7G basis has been shown as a function of $1/k$, where $k$ is the $k$-th coordination sphere included in the procedure described in Sec.~\ref{sec:level_3.2}. For $k\geq6$ we have already very reliable results. Second, the nearest-neighbor overlap $S$, justifying the tight--binding approximation, has been shown in Fig.~\ref{fig:S_vs_R_CH}. We see that for $R/a_0 > 2$ the overlap $S$ is $<1/2$. The difference between the STO-3G and STO-7G is small in that respect. These factors determine reliability of our results. Finally, in Fig.~\ref{fig:wanniers} we have plotted the nearest neighboring optimized Wannier functions when taking STO-7G Gaussian basis. The characteristic cusps appearing at the lattice position have been smeared out to some extent. One should note that the functions contract by up 30\% for a small lattice parameter. This is the reason why the inclusion of the electronic correlations extends the regime of applicability of the tight-binding approximation. One very important feature in Fig.~\ref{fig:wanniers} should be emphasized. Namely, the wave functions are almost the same independently of the method chosen to calculate them. This similarity explains why the microscopic parameters listed in Table I are close in value. Hence, it is the correlation energy which differentiates between the methods.

One should also mention that the GWF solution requires determination of the statistical distribution function $n_k(g)$ ($n_k(g)\equiv n_{k\uparrow}=n_{k\downarrow}$ for paramagnetic, half-filled band case), which is contained in expression ~(\ref{eq:33}). This question has been analyzed in detail elsewhere \cite{JK_JS_WW_APP}. The distribution function differs from that obtained exactly for nanoscopic systems \cite{Spalek_Podsiadly}. Namely, in the latter case the statistical distribution $n_k$ evolves from a Fermi-Dirac-momentum to continuous momentum distribution, without a characteristic cusp at Fermi momentum discussed by Metzner and Vollhardt \cite{Metzner_Vollhardt1}.

Strictly speaking, the expressions (\ref{eq:32})-(\ref{eq:34}) for the ground state energy are functionals of the wave functions $\{w_i(\textbf{r})\}$. So, to obtain the renormalized wave functions one should solve the Euler- Schr\"odinger equation which takes the form
\begin{equation}
\frac{{\delta E_G }}{{\delta w_i^* ({\bf{r}})}} - \nabla  \cdot \frac{{\delta E_G }}{{\delta \nabla w_i^* ({\bf{r}})}} = \lambda w_i(\textbf{r}),
\label{eq:35}
\end{equation}
where $\lambda$ is the renormalized single-particle energy.
This equation is very difficult to solve when e.g. (\ref{eq:18})
 is taken for $E_G$. This is a nonlinear, nonlocal differential-integral equation. To simplify the situation, we have chosen the trial wave functions with the variational parameter being the size of the selected Slater or Gaussian starting atomic wave function.
 
In the next two sections we apply the same type of approach to higher-dimension lattices. Only GA is considered as the other two approaches are possible in the $D=1$ case only.

\section{\label{sec:level_5}Two dimensions\protect}
For the lattices of the dimension $D>1$ there is no exact analytical solution of Hubbard model, except for the solution in $D=\infty$ limit \cite{Metzner_Vollhardt2}. Therefore, to illustrate our method for extended systems for $\infty>D>1$ we use the GA solution \cite{Gutzwiller, Spalek_Oles_Honig, Ogawa_Kanda_Matsubara}. Within this scheme, the ground state energy can be calculated explicit and has the form
\begin{equation}
\frac{{E_G }}{N} = \varepsilon _a^{eff}  - |\bar \varepsilon |\left[ {1 - \frac{U}{{U_c }}} \right]^2,
\label{eq:36}
\end{equation}
where $U_c  = 8|\bar \varepsilon |$ is the critical value, for which we have a mean-field metal-insulator transition and $\bar \varepsilon $ is the average bare band energy. The $\bar \varepsilon $ is determined here from the expression, which in the paramagnetic case amounts to 
\begin{equation}
\bar \varepsilon  = N^{ - 1} \left\langle {\Psi _0 } \right|\sum\limits_{ij\sigma } {t_{ij} } a_{i\sigma }^ +  a_{j\sigma } \left| {\Psi _0 } \right\rangle  = \frac{2}{N}\sum\limits_{|{\bf{k}}| < k_F } {\varepsilon ({\bf{k}}) < 0},
\label{eq:37}
\end{equation}
with
\begin{equation}
\varepsilon ({\bf{k}}) = \frac{1}{{N}}\sum\limits_{ij} {t_{ij} e^{ i{\bf{k}}({\bf{R}}_i  - {\bf{R}}_j )} }
\label{eq:38}
\end{equation}
being the bare band energy of particle with quasimomentum $\textbf{k}$. In thermodynamical limit, $\bar \varepsilon$ can be calculated through the integral
\begin{equation}
\bar \varepsilon  = 2/V^* \int\limits_{V_F } \varepsilon ({\bf{k}})d^D {\bf{k}}=2\int\limits_{\varepsilon ({\bf{k}})_{min}}^{\varepsilon _F}{d \varepsilon \rho(\varepsilon) \varepsilon},
\label{eq:39}
\end{equation}
where $V^*$ is the volume of the primitive unit cell in $\bf{k}$ space, and $V_F$ represents the volume encompassed by the bare Fermi surface $S_F$. The second integral represents the integration over the density of states per atom spin (DOS) $\rho(\varepsilon)$ and with $\varepsilon_F$ being the Fermi energy

The numerical method of determining DOS was that of Buchheit and Loly \cite{Buchheit_Loly}. In the original work, the authors divided band into 100 intervals and the number of generated $\bf{k}$ was $3\cdot 10^5$. In this work, we divide the band into 500 intervals with the $10^9$ generated wave vectors, which reproduce well the known analytic results for the chain. The average bare band energy per site obtained with the help of thus generated DOS is displayed for selected lattices in Table~\ref{tab:tab_eps_sr}.

\begin{table}
\caption{\label{tab:tab_eps_sr}Average bare band energies (per site) for the lattices analyzed in this paper for the half-filled band.}
\begin{tabular}{lc}
\hline \hline
linear chain (CH)&$\bar \varepsilon  =  - (4/\pi)\:t $ \\
square lattice (SQ)&$\bar \varepsilon  =  - 1.62\:t $ \\
triangular lattice (TR)&$\bar \varepsilon  =  - 1.98\:t $\\
simple cubic (SC)&$\bar \varepsilon  =  - 2.01\:t$\\
body center cubic (BCC)&$\bar \varepsilon  =  - 2.16\:t$\\
face center cubic (FCC)&$\bar \varepsilon  =  - 2.64\:t$\\
\hline \hline
\end{tabular}
\end{table}

In this Section we consider the square (SQ) and the triangular (TR) lattices. The single-particle attractive Coulomb potential was constructed for $j\in S_k(i) \cup S_k(i+1)$ composed of $k=13$ spheres of coordination for SQ and $k=11$ for TR cases respectively. In both situations such choice corresponds to up to five lattice parameters and involves 92 atomic sites for SQ and 102 atomic sites for TR cases. The comparison of the $R$ dependence of selected microscopic parameters is provided in Table~\ref{tab:params_2D}. Critical lattice parameter for metal-insulator transition (MIT) is $R_c\approx 3.648\:a_0 \approx1.930\:$\AA\ for SQ and $R_c\approx 3.985\:a_0 \approx2.109\:$\AA\ for TR. We remind the reader, that the corresponding quantity for the Hubbard chain in GA is $R_c\approx 3.318\:a_0 \approx1.756\:$\AA.\ The differences are caused by the different n.n number $z=2, 4$ and $6$, respectively.

\begin{table}
\caption{\label{tab:params_2D}Selected microscopic parameters and quantities vs. lattice spacing for SQ and TR lattices within STO-3G basis, using the GA expression for the ground state energy. The critical distance $R_c/a_0$ for the mean field metal-insulator transition is also specified.}
\begin{tabular}{c|cc||cc||cc||cc||cc}
\hline \hline 
  & SQ & TR & SQ & TR & SQ & TR & SQ & TR & SQ & TR	\\
\cline{2-11}  
$R/a_0$& \multicolumn{2}{c||}{$\varepsilon _a^{eff}$ (Ry)}& \multicolumn{2}{c||}{ $-t$ (Ry)}& \multicolumn{2}{c||}{$U$ (Ry)}& \multicolumn{2}{c||}{$K$ (Ry)}& \multicolumn{2}{c}{$E_G$ (Ry)}\\
\hline
2.0&-0.135&-0.034&-0.494&-0.357&1.938&2.087&0.939&0.946&-0.524&-0.315\\
2.5&-0.606&-0.549&-0.285&-0.219&1.701&1.784&0.769&0.770&-0.741&-0.651\\
3.0&-0.808&-0.777&-0.181&-0.144&1.527&1.579&0.650&0.649&-0.844&-0.804\\
3.5&-0.902&-0.884&-0.121&-0.100&1.418&1.446&0.563&0.562&-0.904&-0.886\\
$(R/a_0)_c$&-0.906\footnotemark[1]&-0.936\footnotemark[2]&-0.090\footnotemark[1]&-0.072\footnotemark[2]&1.398\footnotemark[1]&1.416\footnotemark[2]&0.542\footnotemark[1]&0.538\footnotemark[2]&-0.919\footnotemark[1]&-0.906\footnotemark[2]\\
\hline
$R/a_0$& \multicolumn{2}{c||}{$\beta$}& \multicolumn{2}{c||}{ $\gamma$}& \multicolumn{2}{c||}{$\alpha$}& \multicolumn{2}{c||}{$S$}& \multicolumn{2}{c}{$d$}\\
\hline
2.0&1.417&1.322&0.295&0.186&1.233&1.384&0.466&0.395&0.174&0.158\\
2.5&1.221&1.207&0.205&0.146&1.174&1.241&0.358&0.323&0.135&0.121\\
3.0&1.129&1.135&0.153&0.117&1.107&1.139&0.281&0.264&0.087&0.077\\
3.5&1.071&1.083&0.111&0.091&1.067&1.078&0.212&0.207&0.023&0.022\\
$(R/a_0)_c$&1.057\footnotemark[1]&1.068\footnotemark[2]&0.099\footnotemark[1]&0.082\footnotemark[2]&1.063\footnotemark[1]&1.067\footnotemark[2]&0.192\footnotemark[1]&0.187\footnotemark[2]&0.000\footnotemark[1]&0.000\footnotemark[2]\\
\hline \hline
\end{tabular}
\footnotetext[1]{The value $(R/a_0)_c \approx 3.648$ for SQ.}
\footnotetext[2]{The value $(R/a_0)_c \approx 3.668$ for TR.}
\end{table}

\begin{figure}
\centerline{
\includegraphics[width=10.5cm]{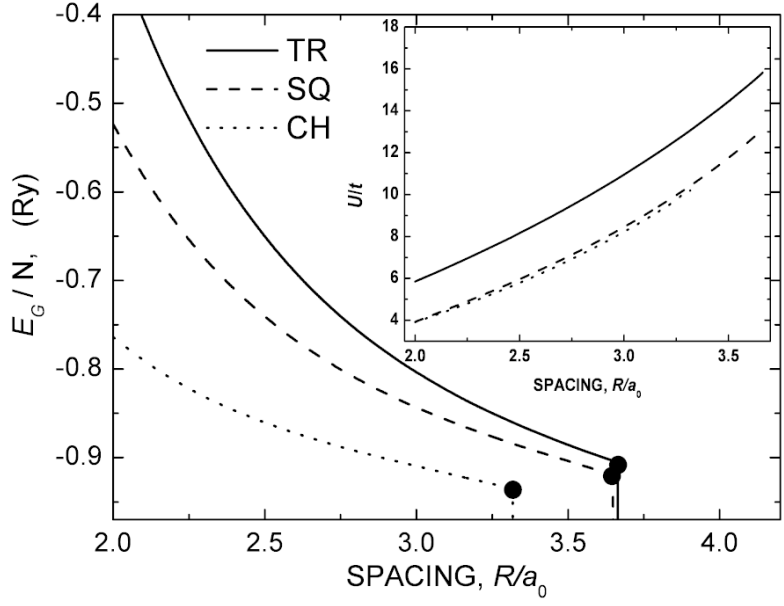}}
\caption{\label{fig:EG_2D}Ground state energy vs. $R$ for linear chain (CH) square lattice (SQ), and triangular lattice (TR). Inset: $U/t$ ratio as a function of lattice parameter. The solid circles mark the MIT instability within GA solution.}
\end{figure}

\begin{figure}
\centerline{
\includegraphics[width=10.5cm]{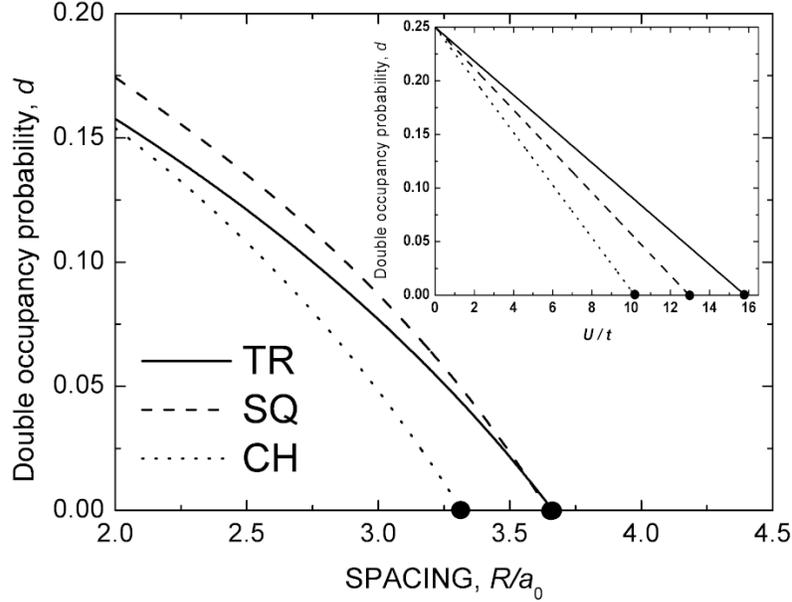}}
\caption{\label{fig:d_2D}Double occupancy probability $d= \langle n_{i\uparrow}n_{i\downarrow}\rangle$ vs. $R$  for the three low dimensional lattices: linear chain, square- and triangular-lattices. Inset: same as a function of $U/t$. The solid points mark the critical distances for MIT within GA.}
\end{figure}

The ground state energy $E_G /N$ vs. $R$ for both lattices is shown in Fig.~\ref{fig:EG_2D}. We see similarity for TR and SQ lattices, even though the $U/t$ ratio similar for CH and SQ (cf. inset). The differences are included by different dependence of the double occupancy, as shown in Fig.~\ref{fig:d_2D}. The probability $d$ is analytical function of $U/t$, namely
\begin{equation}
d = \frac{1}{4}\left( {1 - \frac{U}{{8\left|\bar \varepsilon \right| }}} \right).
\label{eq:42}
\end{equation}

\begin{figure}
\centerline{
\includegraphics[width=10.5cm]{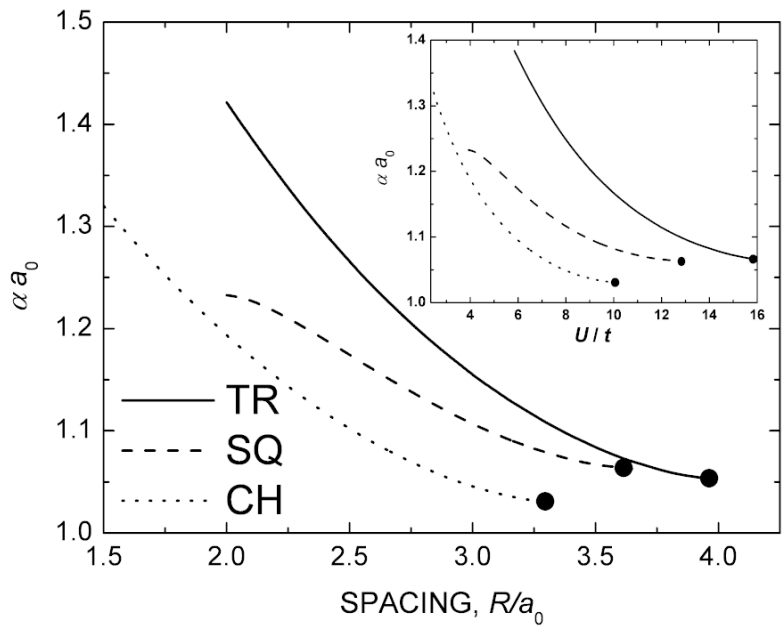}}
\caption{\label{fig:alfa_2D}Inverse orbital size $\alpha$ vs. $R$ of the atomic-like (Gaussian) wave function for the lattices specified. Inset: Same as a function of $U/t$. Note that the intersite interation is calculated in the Hartree-Fock approximation in the metallic state.}
\end{figure}

The interesting quantity in this case is also the size $a=1/\alpha$ of the single atomic wave function composing the Wannier function. This quantity is plotted as a function of lattice parameter in Fig.~\ref{fig:alfa_2D}. In all cases, correlations diminish the size of the starting atomic-like wave function. The size renormalization is 25-40\% reduction for small lattice parameter. This result is counterintuitive, as one would expect, that the Coulomb repulsion increase the atomic size. This is not the case and, parenthetically, also the reason why the He atom, not H, is the smallest atom in the Universe \cite{Bethe_Salpeter}.

\section{\label{sec:level_6}Three dimensions\protect}
In this Section we discuss the three cubic lattices. As in the case of 2D lattices, we select the sites generating the single-particle potential surrounding the central atomic site and one of its neighbors by atomic potentials distant by up to five lattice parameter. This case corresponds to the set of sites $j \in S_k (i) \cup S_k (i + 1)$ with $k=22$ (596 lattice sites) for SC structure, $k=35$ (1026 sites) for BCC structure, and $k=47$ (2344 sites) for FCC lattice.

Selected parameters of Hubbard Hamiltonian are shown in Table~\ref{tab:param_3D}. Note, that now $R$ is the distance between the nearest neighbors, not the lattice parameter.

\begin{table*}
\caption{\label{tab:param_3D}Selected microscopic parameters for cubic lattices vs. $R$.}
\begin{tabular}{c|ccc||ccc||ccc||ccc}
\hline \hline 
  & SC & BCC & FCC & SC & BCC & FCC & SC & BCC & FCC & SC & BCC & FCC\\
\cline{2-13}  
$R/a_0$ &&$\varepsilon _a^{eff}$ (Ry)& & & $-t$ (Ry)& & &$U$ (Ry)& & &$K$ (Ry)\\
\hline
3.3&-0.718&&-0.652&-0.222&&-0.167&1.502&&1.621&0.599&&0.598\\
3.4&-0.764&&-0.699&-0.197&&-0.151&1.490&&1.601&0.582&&0.582\\
3.5&-0.799&&-0.738&-0.177&&-0.138&1.474&&1.579&0.567&&0.566\\
3.6&-0.828&&-0.770&-0.161&&-0.127&1.459&&1.557&0.551&&0.551\\
3.7&-0.852&&-0.799&-0.146&&-0.117&1.444&&1.536&0.537&&0.537\\
3.8&-0.872&&-0.824&-0.133&&-0.108&1.432&&1.516&0.524&&0.523\\
3.9&-0.890&-0.845&-0.845&-0.121&-0.145&-0.100&1.422&1.386&1.498&0.511&0.509&0.510\\
4.0&-0.905&-0.872&-0.864&-0.110&-0.126&-0.093&1.415&1.396&1.482&0.498&0.497&0.498\\
4.1&-0.918&-0.893&-0.881&-0.100&-0.112&-0.086&1.413&1.399&1.469&0.487&0.486&0.486\\
4.2&-0.929&-0.910&-0.895&-0.090&-0.100&-0.079&1.417&1.404&1.459&0.475&0.475&0.475\\
4.3&-0.930\footnotemark[1]&-0.924&-0.908&-0.088\footnotemark[1]&-0.088&-0.073&1.419\footnotemark[1]&1.415&1.452&0.473
\footnotemark[1]&0.464&0.464\\
$(R/a_0)_c$&&-0.929\footnotemark[2]&-0.915\footnotemark[3]&&0.083\footnotemark[2]&-0.069\footnotemark[3]&&1.426\footnotemark[2]&1.451\footnotemark[3]&&0.459\footnotemark[2]&0.457\footnotemark[3]\\
\hline \hline
$R/a_0$&& $\beta$ & & & $\gamma$ & & & $\alpha\:a_0$ & & & $S$\\
\hline
3.3&1.137&&1.124&0.126&&0.074&1.072&&1.165&0.243&&0.196\\
3.4&1.112&&1.106&0.113&&0.069&1.082&&1.165&0.221&&0.180\\
3.5&1.095&&1.092&0.104&&0.064&1.084&&1.160&0.204&&0.167\\
3.6&1.081&&1.081&0.096&&0.060&1.084&&1.154&0.188&&0.156\\
3.7&1.069&&1.071&0.088&&0.057&1.083&&1.147&0.174&&0.145\\
3.8&1.058&&1.062&0.081&&0.053&1.083&&1.139&0.160&&0.135\\
3.9&1.049&1.082&1.055&0.074&0.079&0.050&1.084&1.026&1.133&0.147&0.175&0.126\\
4.0&1.040&1.060&1.048&0.067&0.068&0.047&1.087&1.052&1.128&0.134&0.149&0.117\\
4.1&1.032&1.047&1.041&0.060&0.060&0.044&1.093&1.067&1.125&0.120&0.131&0.107\\
4.2&1.025&1.036&1.034&0.053&0.053&0.040&1.104&1.081&1.124&0.105&0.114&0.098\\
4.3&1.024\footnotemark[1]&1.026&1.028&0.051\footnotemark[1]&0.045&0.037&1.107\footnotemark[1]&1.100&1.126&0.103\footnotemark[1]&0.097&0.088\\
$(R/a_0)_c$&&1.021\footnotemark[2]&1.025\footnotemark[3]&&0.041\footnotemark[2]&0.034\footnotemark[3]&&1.114\footnotemark[2]&1.129\footnotemark[3]&&0.088\footnotemark[2]&0.082\footnotemark[3]\\
\hline \hline
\end{tabular}
\footnotetext[1]{The value $(R/a_0)_c \approx 4.216$ for SC.}
\footnotetext[2]{The value $(R/a_0)_c \approx 4.352$ for BCC.}
\footnotetext[3]{The value $(R/a_0)_c \approx 4.366$ for FCC.}
\end{table*}

\begin{figure}
\centerline{
\includegraphics[width=10.5cm]{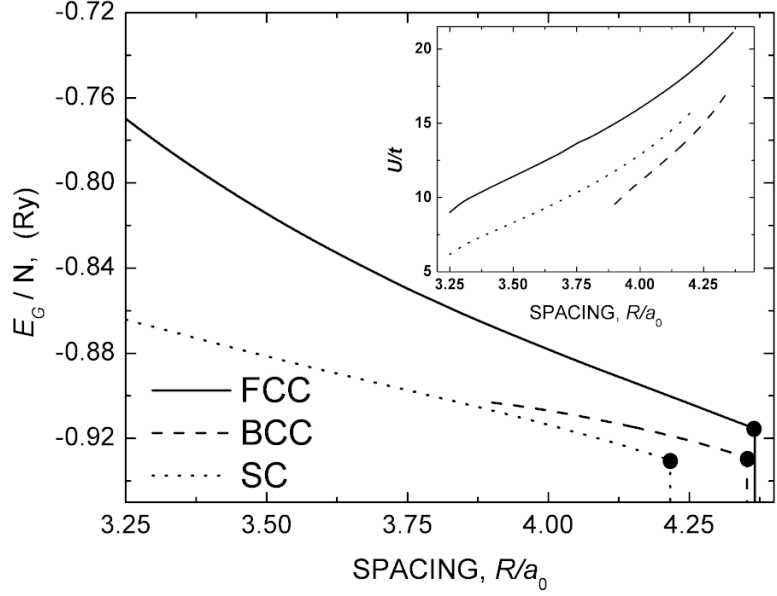}}
\caption{\label{fig:EG_3D}Ground state energy vs. the nearest-neighbor distance $R$ for the cubic lattices. Inset: $U/t$ as a function of $R$ for those cases. The solid circles mark the MIT instability.}
\end{figure}

\begin{figure}
\centerline{
\includegraphics[width=10.5cm]{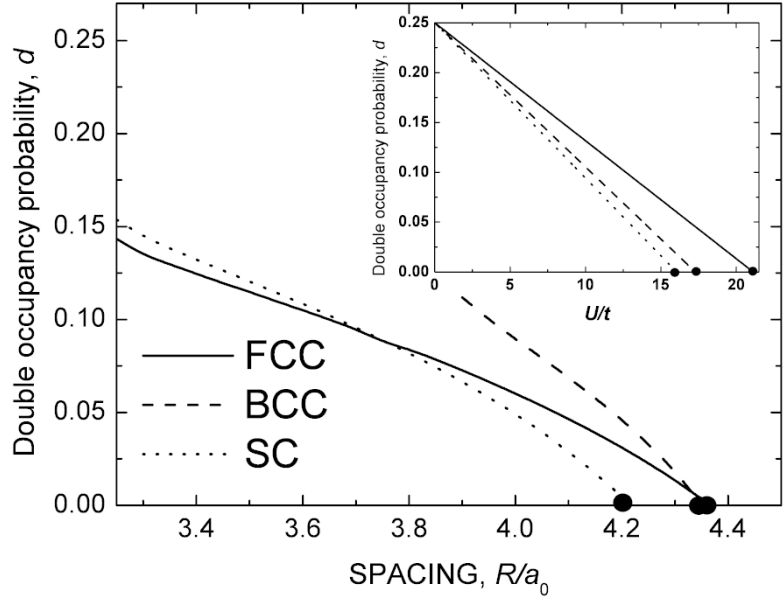}}
\caption{\label{fig:d_3D}Double occupancy probability $d= \langle n_{i\uparrow}n_{i\downarrow}\rangle$ as a function of the distance $R$ between nearest neighbors for the cubic lattices. Inset: Double occupancy probability vs. $U/t$.}
\end{figure}

\begin{figure}
\centerline{
\includegraphics[width=10.5cm]{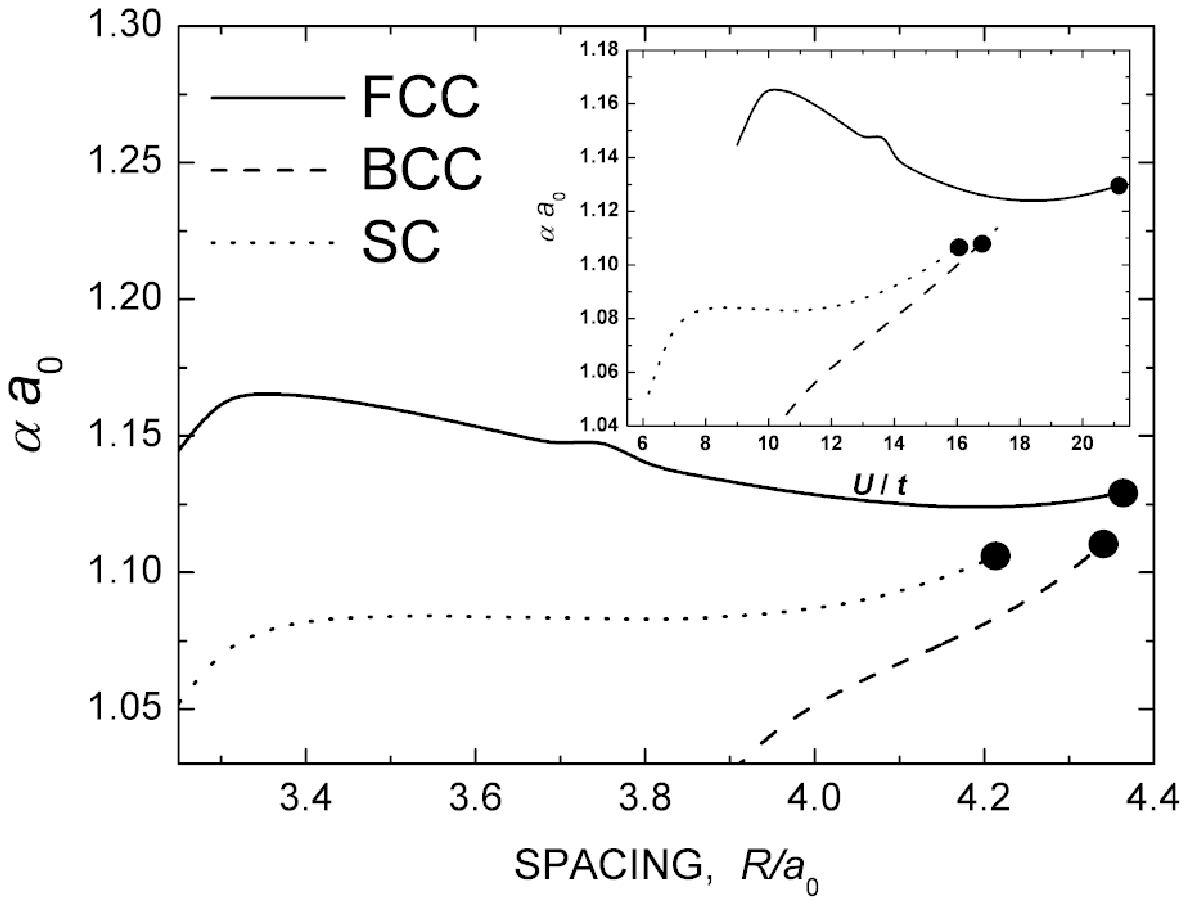}}
\caption{\label{fig:alfa_3D}Inverse size $\alpha$ vs. $R$ of the atomic-like (Gaussian) wave function for the lattices specified. Inset: Same as a function of $U/t$.}
\end{figure}

In Fig.~\ref{fig:EG_3D} we present the ground state energy vs. $R$ As before, the Hubbard cubic lattice of 1s are not stable on absolute energy scale. The double occupancy probability displayed in Fig.~\ref{fig:d_3D}. 
The MIT transition at $T=0$ (from the paramagnetic metal to paramagnetic insulator) is continuous. For the sake of completeness, in Fig.~\ref{fig:alfa_3D} we have also shown the inverse size of atomic wave function vs. nearest-neighbor distance $R$.

\begin{figure*}
\centerline{
\includegraphics[width=15cm]{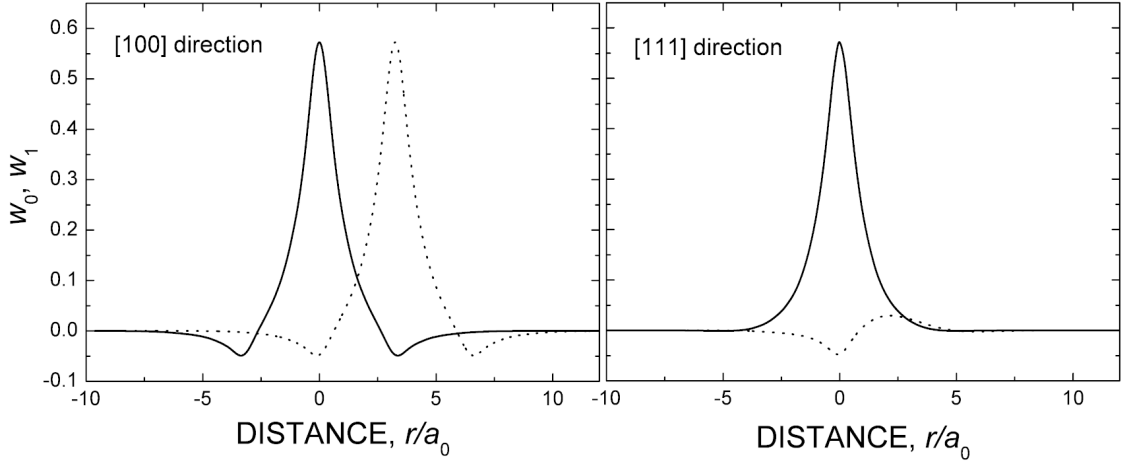}}
\caption{\label{fig:Wanniers_SC}The n.n. renormalized Wannier functions along [100] direction for SC structure (left); right: the functions along [111] direction.}
\end{figure*}

In Figs.~\ref{fig:Wanniers_SC} we draw the nearest neighboring Wannier functions for SC structure along the directions [100] (left) and [111] (right) locating one of then at the origin. We see that overlap (and hopping) integral are strongly dependent on the direction, as one would expect. A similar effect is expected for the case of square lattice for $D=2$.

\section{\label{sec:level_7}Discussion and Outlook}
In this paper we have shown on the example of a simple workable scheme - the Gutzwiller ansatz for the extended Hubbard model - how the calculation of single-particle wave function \textit{in the correlated state} can be incorporated within the theory of correlated fermions in dimmensions $D=1,2,3$. What is even more important, the dependence of the physical properties (e.g. the ground state energy) is calculated as a function of interatomic distance and this dependence is not related linearly to that as a function of the parameter $U/t$. In other words, only the explicit dependence on $R$ provides the proper scaling of the quantities. Additionally, the Wannier orbitals calculated from the variational principle for the energy of the correlated-electron state are renormalized in the sense that they are more tightly bound to their parent ions than their bare correspondants, thus extending the applicability of the tight-binding approach for narrow-band systems. These are the most important qualitative conclusions coming from our method of approach. In essence, as for the correlated system the interaction is comparable (or even stronger) than the single-particle part, the method reverts the usual approach, i.e. we determine the single-particle wave function only after diagonalization of the many-particle parametrized Hamiltonian. Such a method is fully implemented here, albeit for the case of a narrow s-band.

The importance of the Hubbard model in its present form for the real 3d systems such as oxides, manganites or intermetallic compounds, is limited. This is because the orbital degenerancy of the 3d states is not included explicitly and the interorbital Hunds's rule and direct Coulomb interactions do not appear here. \textit{Therefore, the factors such as the Hund's rule are neglected}. It would be important to extend there present method by including crystal-field effects due to anions and interatomic interorbital interactions. The first steps in this direction have been undertaken already \cite{GorlichPhD}. The same remark concerns the inclusion of the 3d-2p hybrydization in the metallic oxides and other related compounds. On the example of our model calculations we see that while the single-particle wave function does not depend strongly on the method of treating the many-particle Hamiltonian in the Fock space, (cf. Sec. IVB), the two-particle correlation functions (e.g. $d(R)$), are strongly method-dependent. This is the reason why many-body renormalizations of the quasiparticle characteristics are not computable in a reliable manner within any of the present band-structure calculations. So, extensions of this method to the degenerate Hubbard model solution combined with wave-function adjustment in the correlated state, is highly desirable.

\section{\label{sec:level_8}Acknowledgement}
We thank Dr. Adam Rycerz for discusions on the Gaussian basis, Maciej Ma\'{s}ka and Krzysztof Ro\'{s}ciszewski for remarks on detailed points concerning various aspects of this work. The work was supported by the Ministry of Science and Higher Education, Grant No. 1P03B 001 29. This work represent a part of the Ph.\;D.\;Thesis of one of the authors (J.K.) submitted to the Faculty of Physics, Astronomy, and Applied Computer Science of Jagiellonian University \cite{KurzykPhD}. 
The work was performed in part under the auspices of the \textit{Marek Kac Interdisciplinary Center for Complex Systems} (COCOS) financed by the Marie Curie TOK European Union, Grant No. HTDK-CT-2004-517186.\\

\appendix

\section{The expressions for coefficients of $A$ and $B$}
Below we list the explicit expressions for the coefficients $A$ and $B$ defined in Sec.~\ref{sec:level_3.2}
\begin{subequations}
for the linear chain (CH)
\begin{equation}
A = 1 + S_2 ,\ \ \ \ B = 3 + S_3 ,
\end{equation}
the square lattice (SQ)
\begin{equation}
A = 1 + 2S_2  + S_3 ,\ \ \ \ B = 9 + 4S_4  + S_6 ,
\end{equation}
the triangular lattice (TR)
\begin{eqnarray}
A &=& 1 + 2S_1  + 2S_2  + S_3 ,\nonumber \\
B &=& 1 + 15S_1  + 6S_2  + 6S_3  + 6S_4  + S_5 ,
\end{eqnarray}
the simple cubic (SC)
\begin{equation}
A = 1 + 4S_2  + S_4 ,\ \ \ \ B = 15 + 8S_3  + 12S_5  + S_8,
\end{equation}
the body-centered cubic (BCC)
\begin{eqnarray}
A &=& 1 + 3S_2  + 3S_3  + S_5 ,\nonumber \\
B &=& 27S_1  + 27S_4  + 9S_7  + S_{10}
\end{eqnarray}
and the face-centered cubic (FCC)
\begin{eqnarray}
A &=& 1 + 4S_1  + 2S_2  + 4S_3  + S_4 ,\nonumber \\
B &=& 4 + 45S_1  + 12S_2  + 36S_3  + 12S_4 + 18S_5  + 4S_6  + 12S_7  + S_9.
\end{eqnarray}
\end{subequations}
These lattices coefficients are used in the subsequent analysis.

\section{The expressions for coefficients of $\varepsilon _a$ and $t$ for the all lattices under consideration}

For the single-particle basis defined in subsection ~\ref{sec:level_3.2}, the parameters $\varepsilon _a$ and $t$ acquire the form for a linear chain 
\begin{subequations}
\begin{eqnarray}
\varepsilon _a &=& \beta ^2 T_0  - 4\beta \gamma T_1  + 2\gamma ^2 (T_0  + T_2 ), \\
-t \equiv t_{i,i + 1} &=& \beta ^2 T_1  - 2\beta \gamma (T_0  + T_2 ) + \gamma ^2 (3T_1  + T_3 ).\ \ \  \ \ \ \label{eq:CH}
\end{eqnarray}
\end{subequations}

For the square lattice (SQ) the single parameters are expressed through their atomic (Gaussian) basis correspondents as follows

\begin{subequations}
\begin{eqnarray}
\varepsilon _a  &=& \beta ^2 T_0  - 8\beta \gamma T_1  + 4\gamma ^2 (T_0  + 2T_2  + T_3 ),\\
-t &=& \beta ^2 T_1  - 2\beta \gamma (T_0  + 2T_2  + T_3 ) + \gamma ^2 (9T_1  + 6T_4  + T_6 ).
\label{eq:eps_t_SQ}
\end{eqnarray}
\end{subequations}
For the triangular lattice they are
\begin{subequations}
\begin{eqnarray}
\varepsilon _a &=& \beta T_0  - 12\beta \gamma T_1  + 6\gamma ^2 (T_0  + 2T_1  + 2T_2  + T_3 ),\ \ \ \ \ \ \\
-t &=& \beta ^2 T_1  - 2\beta \gamma (T_0  + 2T_1  + 2T_2  + T_3 )+ \gamma ^2 (2T_0  + 15T_1  + 6T_2  + 6T_3  + 6T_4  + T_5 ).\ \ \ \ \ \
\label{eq:eps_t_TR}
\end{eqnarray}
\end{subequations}

For the cubic lattices the single particle parameters are expressed through their atomic (Gaussian) basis correspondents as follows:\\
for SC
\begin{subequations}
\begin{eqnarray}
\varepsilon _a  &=& \beta ^2 T_0  - 12\beta \gamma T_1  + 6\gamma ^2 (T_0  + 4T_2  + T_4 ),\ \ \ \\\
-t &=& \beta ^2 T_1  - 2\beta \gamma (T_0  + 4T_2  + T_4 ) + \gamma ^2 (15T_1  + 8T_3  + 12T_5  + T_8 ),
\label{eq:eps_t_SC}
\end{eqnarray}
\end{subequations}
for BCC
\begin{subequations}
\begin{eqnarray}
\varepsilon _a  &=& \beta ^2 T_0  - 16\beta \gamma T_1  + 8\gamma ^2 (T_0  + 3T_2  + 3T_3  + T_5 ),\ \ \ \ \ \ \ \ \ \\\
-t &=& \beta ^2 T_1  - 2\beta \gamma (T_0  + 3T_2  + 3T_3  + T_5 ) + \gamma ^2 (27T_1  + 27T_4  + 9T_7  + T_{10} ),
\label{eq:eps_t_BCC}
\end{eqnarray}
\end{subequations}
and for FCC
\begin{subequations}
\begin{eqnarray}
\varepsilon _a  &=& \beta ^2 T_0  - 24\beta \gamma T_1 + 12\gamma ^2 (T_0  + 4T_1  + 2T_2  + 4T_3  + T_4 ),\\
 -t &=& \beta ^2 T_1  - 2\beta \gamma (T_0  + 4T_1  + 2T_2  + 4T_3  + T_4 )+ \nonumber\\
 &\gamma ^2& (4T_0  + 45T_1  + 12T_2  + 36T_3  + 12T_4  + 18T_5  + 4T_6  + 12T_7  + T_9 ).\ \ \ \ \ \ \ \ \ \ \ 
\label{eq:eps_t_FCC}
\end{eqnarray}
\end{subequations}


\end{document}